\documentclass[11pt,a4paper]{article}
\pdfoutput=1
\usepackage{jheppub}
  
\usepackage{graphicx}
\usepackage{amsmath,amssymb,amsfonts}
\usepackage{verbatim}   
\usepackage{subfigure}  
\usepackage{acronym}
\usepackage{hyperref}
\usepackage{mathrsfs}
\usepackage{multirow}
 \usepackage{slashed}
 \usepackage{url}

\usepackage{amsmath}
\usepackage{amsfonts}
\usepackage{amssymb}

\newcommand{\fb}{{\mathrm {fb}}}
\def\OO{\mathcal{O}}

\title{The impact of heavy-quark loops on LHC dark-matter searches}

\preprint{OUTP-12-15P}

\author[a]{Ulrich Haisch,}
\author[a]{Felix Kahlhoefer}
\author[a,b]{and James Unwin}
\affiliation[a]{Rudolf Peierls Centre for Theoretical Physics, University of Oxford, 1 Keble Road, Oxford OX1 3NP, United Kingdom}
\affiliation[b]{Mathematical Institute, University of Oxford, 24-29 St Giles, Oxford, OX1 3LB, UK}
\emailAdd{u.haisch1@physics.ox.ac.uk}
\emailAdd{felix.kahlhoefer@physics.ox.ac.uk}
\emailAdd{j.unwin1@physics.ox.ac.uk}

\abstract{If only tree-level processes are included in the analysis, LHC monojet searches give weak constraints on the dark matter-proton scattering cross section arising from the exchange of a new heavy scalar or pseudoscalar mediator with Yukawa-like couplings to quarks. In this letter we calculate the constraints on these interactions from the CMS $5.0 \, \fb^{-1}$ and ATLAS $4.7 \, \fb^{-1}$ searches for jets with missing energy including the effects of heavy-quark loops. We find that the inclusion of such contributions leads to a dramatic increase in the predicted cross section and therefore a significant improvement of the bounds from LHC searches. 

}


\date{\today}

\begin{document}

\maketitle

\section{Introduction}

It is a remarkable fact that one of the leading experiments for dark matter (DM) detection is the LHC. Although any DM particles produced at the LHC will escape from the detector unnoticed, we may observe large amounts of missing transverse energy ($\slashed{E}_T$) if a single jet~($j$) is produced in association with a pair of DM particles. The experimental search for such monojet events provides model-independent bounds on the interaction strength of DM with quarks and gluons, constraining the same parameters as direct detection experiments~\cite{Chatrchyan:2012pa,ATLAS:2012ky}. These searches place the leading (and in some cases only) limits on models of DM over certain regions of parameter space. 

\begin{figure*}[!t]
\centering
\includegraphics[height=0.22\textwidth]{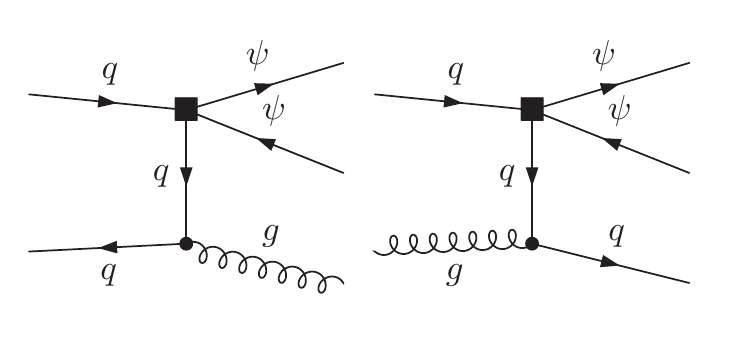}

\includegraphics[height=0.22\textwidth]{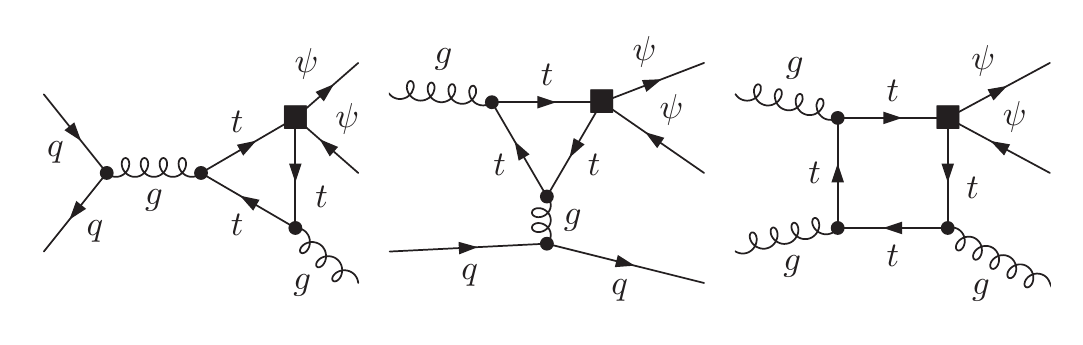}
\vspace{-3mm}
\caption{Typical tree-level (top) and loop-level (bottom) diagrams leading to monojet events. The black squares denote insertions of four-fermion operators.}
\label{Fig1}	
\end{figure*}

If the mediator of the DM interaction is sufficiently heavy, it can be integrated out to obtain an effective higher-dimensional operator that describes the low-energy interactions between DM and Standard Model (SM) states. For instance, the interactions of a fermionic DM particle $\psi$ with SM quarks $q$ via a heavy scalar mediator $\Phi$ with quark couplings proportional to the quark mass $m_q$ can be described by the four-fermion operator
\begin{equation} \label{eq:Opsis}
\mathcal{O}^\psi_s = \frac{m_q}{\Lambda_s^3} \, \bar{q} q \, \bar{\psi} \psi \; ,
\end{equation}
and the DM pair production proceeds through tree-level diagrams like those shown in the upper row of Fig.~\ref{Fig1}. Unfortunately, the suppression scale $\Lambda_s$ in (\ref{eq:Opsis}) is very difficult to constrain with LHC monojet searches \cite{Goodman:2010yf,Bai:2010hh,Goodman:2010ku,Fox:2011fx,Rajaraman:2011wf,Fox:2011pm,Fox:2012ee,MarchRussell:2012hi} since the initial state contains (apart from gluons) only light quarks, which by assumption have small couplings to the scalar mediator~$\Phi$.

In this letter we observe that the situation changes dramatically beyond tree level, since now loop graphs involving virtual top quarks  give rise to a $j + \slashed{E}_T$ signal. Examples of such Feynman diagrams are displayed in the lower row of Fig.~\ref{Fig1}. We find that by including these loop contributions, the predicted cross section for monojet production increases by more than two orders of magnitude leading to bounds on $\Lambda_s$ which are about a factor of 3 stronger than the constraints from the simple tree-level analysis with DM-quark operators. To our knowledge, this important observation has not been made before.

This paper is structured as follows: in Sec.~\ref{Sect2}, we present the various operators which we intend to examine. In Sec.~\ref{Sect3} we calculate the constraints arising from the latest LHC searches for jets with $\slashed{E}_T$, including the full top-quark mass dependence of the squared matrix elements for  DM pair $+$ jet production. Our calculation is performed at the leading order (LO) in QCD, but we will comment on the importance of higher-order effects as well as the applicability of the heavy top-quark approximation. Finally, in  Sec.~\ref{sec:conclusions} we discuss the impact of the enhanced monojet limits in the context of relic density constraints and recent results from direct detection experiments.

\section{Effective operators}
\label{Sect2}

Our focus will be on interactions that result from the exchange of a new heavy scalar or pseudoscalar state which connects SM quarks to DM. In the case that the scalar mediator is a SM singlet it can couple to quarks via mixing with the Higgs and the induced couplings will be proportional to the SM Yukawa couplings:
\begin{equation} \label{eq:calLs}
\mathcal{L}_\Phi = g_q \, \frac{m_q}{v} \, \bar{q} q \, \Phi + g_\psi \, \bar{\psi} \psi \, \Phi \; ,
\end{equation}
where  $v\simeq246 \, {\rm GeV}$ is the Higgs vacuum expectation value. Integrating out the mediator $\Phi$, we obtain the effective interaction (\ref{eq:Opsis}) and the scale $\Lambda_s$ is related to the fundamental couplings by $\Lambda_s^3 = v M_\Phi^2 / (g_q g_\psi)$. If DM is a scalar $\phi$ rather than a fermion  $\psi$, we obtain in complete analogy the effective operator
\begin{equation} \label{eq:O2}
 \OO^\phi_{s} = \frac{m_q}{\Lambda_s^2}\,  \bar q q \, \phi^{\dagger} \phi \; .
\end{equation}

The other type of portal interaction for which the effective operator may naturally have Yukawa-like couplings involves heavy pseudoscalar mediators. The corresponding effective operator is \footnote{For brevity, we omit the case of scalar DM with a pseudoscalar mediator, since this interaction violates CP, making this scenario rather less appealing. Note, however, that the limits for this operator would be comparable to those presented.}
\begin{equation} 
 \OO^\psi_{p} = \frac{m_q}{\Lambda_p^3} \, \bar q \gamma^5 q \, \bar \psi \gamma^5 \psi \; .
\label{eff}
\end{equation}
This case is of particular interest, since direct detection signals associated to $\OO^\psi_{p}$ are spin-dependent and suppressed by powers of $q/m_N \ll 1$, where $q$ is the momentum transfer and $m_N$ is the mass of the target nucleus. Thus, monojet searches provide the only manner to obtain constraints on $\Lambda_p$ for the foreseeable future. 

\section{Limits from monojet searches}
\label{Sect3}

In their most recent analysis with an integrated luminosity of $5.0 \, \fb^{-1}$ at $\sqrt{s}=7 \, {\rm TeV}$ the CMS collaboration~\cite{Chatrchyan:2012pa} found no apparent excesses in their searches for jets with $\slashed{E}_T$ providing the leading monojet bounds on DM. CMS considered events with $\slashed{E}_T>350\,{\rm GeV}$, provided there was a primary jet ($j_1$) with transverse momentum $p_T>110 \, {\rm GeV}$ and pseudorapidity $|\eta|<2.4$. A secondary jet ($j_2$) with $p_T>30 \, {\rm GeV}$ was also permitted if the two jets are not back-to-back: $|\Delta\phi(j_1, j_2)|<2.5$. Events with high-$p_T$ tertiary jets, electrons or muons were vetoed. This null result excludes new contributions to the production cross section in excess of $0.032 \, {\rm pb}$ at $95\%$ confidence level (CL). The ATLAS search~\cite{ATLAS:2012ky} employs very similar cuts and finds a comparable bound on the cross section. We have checked explicitly that using the ATLAS data instead of the CMS data does not modify our results within errors. 

\begin{figure*}[t!]
\centering
\includegraphics[width=0.45\textwidth]{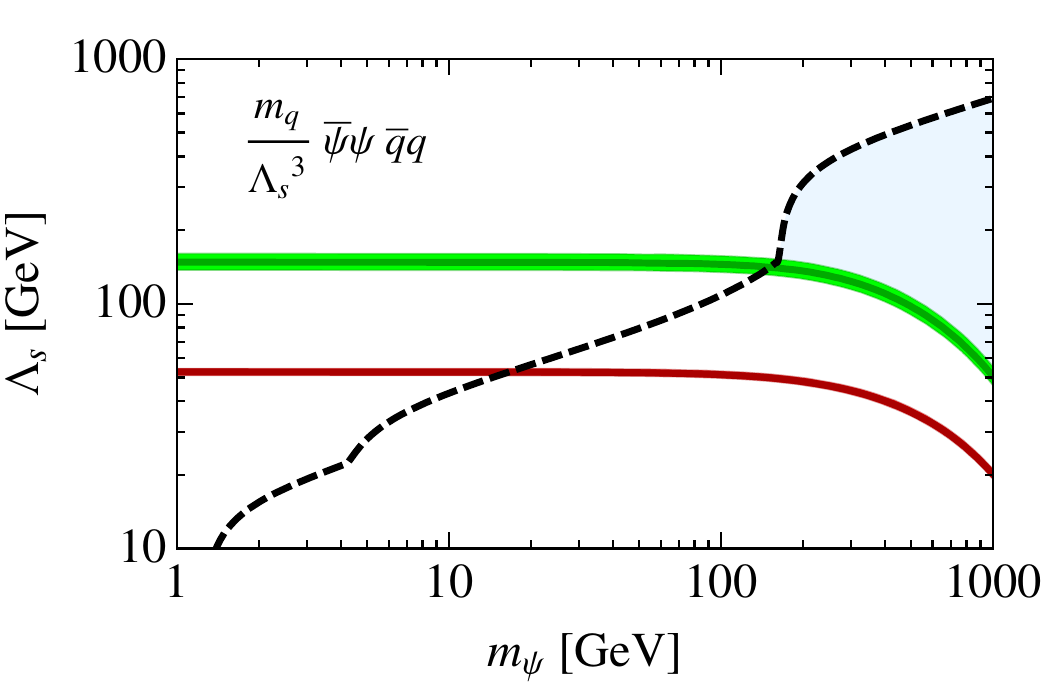}
\quad
\includegraphics[width=0.45\textwidth]{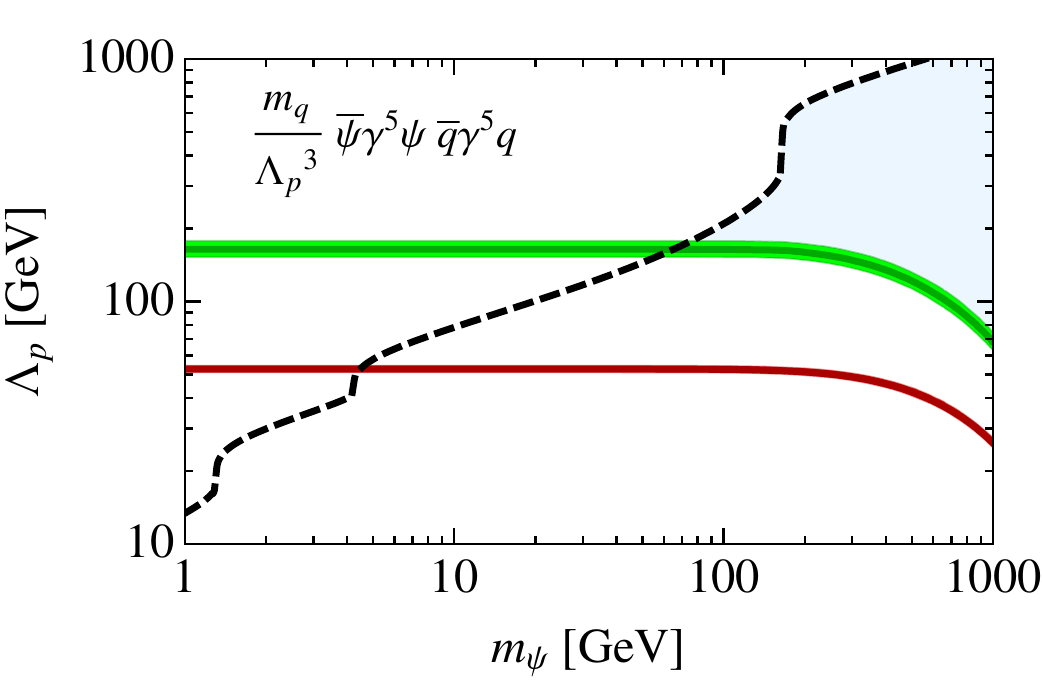}

\includegraphics[width=0.45\textwidth]{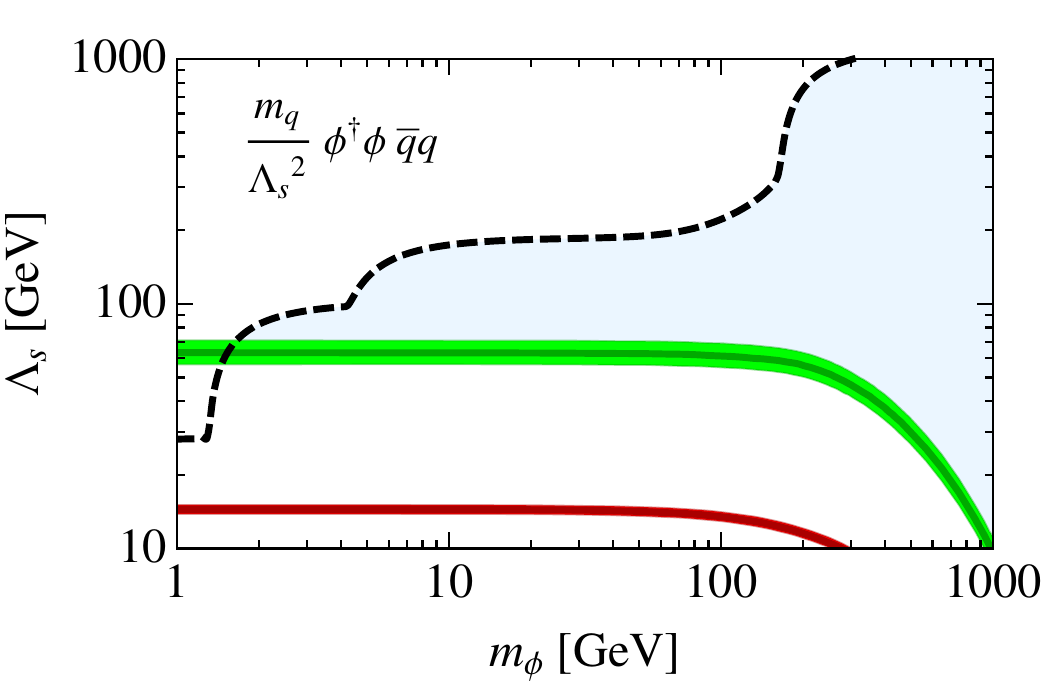}
\caption{\label{Fig2}  LHC monojet bounds on the scale $\Lambda$ at tree-level (red) and loop-level (green) for the effective operators $\mathcal{O}^\psi_s$ (top left), $\mathcal{O}^\psi_p$ (top right), and $\mathcal{O}^\phi_s$ (bottom). The black dashed curves indicate the requirement for the correct relic density. Values of $\Lambda$ above this curve imply an overproduction of DM in the early universe, while values below are not excluded, leading to allowed parameter regions for large DM masses (indicated by a light blue shading). The width of the bands reflect the scale uncertainties. See text for details.} 
 \end{figure*}


To calculate the predicted monojet cross section for the operators introduced above we have implemented each of them in {\tt FeynArts}~\cite{Hahn:2000kx} and  performed the computations with  {\tt  FormCalc} and {\tt   LoopTools}~\cite{Hahn:1998yk}. Furthermore, as an independent cross-check, we have verified our findings with { \tt MCFM}~\cite{MCFM}, modifying the process $p + p \rightarrow H \hspace{0.5mm} (A) + j \rightarrow \tau^+ \tau^- + j$, which is based on the analytical results of~\cite{Ellis:1987xu} for the scalar Higgs case and \cite{Spira:1995rr} for the pseudoscalar Higgs case. Both computations utilise {\tt MSTW2008LO} parton distributions \cite{Martin:2009iq}. In our calculations we do not consider the effects of parton showering and hadronisation or the contribution of additional jets.  As discussed in \cite{Bai:2010hh,Choudalakis:2011bf}, the  first two simplifications are justified, because the primary jet has sufficiently high  $p_T$,  which renders the impact of non-hard QCD radiation small. We will comment on the effect of two-jet events below.

Our results are summarised in Tab.~\ref{table} and displayed in Fig.~\ref{Fig2} for the case where DM is either a Dirac fermion or a complex scalar. For Majorana fermion or real scalar DM, the predicted cross sections are larger by a factor of 2 and so the bounds on $\Lambda$ are stronger by roughly $12\%$. For the scalar (pseudoscalar) operators we find that including the loop-level processes in the calculation increases the predicted monojet cross  sections by a factor of around 500 (900), the precise value depending on the DM mass and the choices of renormalisation ($\mu_R$) and factorisation ($\mu_F$) scales. These numbers translate into an increase of the bounds on $\Lambda$ by a factor of 2.8 (3.1). Note that the limit on $\Lambda_s$ which we obtain from the tree-level processes alone is stronger than the one reported in~\cite{ATLAS:2012ky} by about $25\%$, because we include bottom quarks in the initial state, which we find to give the dominant contribution at tree level.

Before examining the impact of additional QCD radiation, we first discuss whether the above results could have been obtained without performing an actual loop calculation, but rather by employing the heavy top-quark mass limit. For the operator $\mathcal{O}^\psi_{s}$, the effect of heavy-quark loops can be described in this approximation in terms of the effective operator
\begin{equation} \label{eq:gluons}
\mathcal{O}^\psi_{sg}  = \frac{\alpha_s}{4 \Lambda_g^3} \, G^a_{\mu \nu} G^{a\,\mu\nu} \, \bar{\psi} \psi \; .
\end{equation}
The operator induced by $\mathcal{O}^\phi_s$ ($\mathcal{O}^\psi_p$) is obtained by replacing $\bar{\psi} \psi$ with $\phi^\dagger \phi$ ($G^a_{\mu \nu} G^{a\,\mu\nu}$ with $G^a_{\mu \nu} \tilde{G}^{a\,\mu\nu}$).
Effective interactions like  (\ref{eq:gluons}) have been studied previously in the context of monojet searches~\cite{Goodman:2010yf,Goodman:2010ku,Fox:2011pm,Fox:2012ee}. In fact, for $m_t \to \infty$ bounds on these operators can be translated into limits on $\mathcal{O}^{\psi,\phi}_{s}$ and $\mathcal{O}^\psi_{p}$ by the  simple identifications $\Lambda_s = \Lambda_g /(3 \pi)^{1/3}$ and $\Lambda_p = \Lambda_g /(2 \pi)^{1/3}$.

\begin{table}[t!]
\setlength{\tabcolsep}{5pt}
\renewcommand{\arraystretch}{1.2}
\center
\begin{tabular}{|c|c|c|c|} 
\hline
$m_\text{DM}$ [GeV] & $\Lambda^\psi_s$  [GeV] & $\Lambda^\psi_p$  [GeV] & $\Lambda^\phi_s$  [GeV] \\
\hline  \hline 
10 & $148^{+12}_{-11}$ & $164^{+14}_{-11}$ & $63^{+8}_{-6}$\\
\hline
100 & $145^{+12}_{-10}$ & $164^{+14}_{-12}$ &$61^{+8}_{-6}$ \\
\hline
200 & $136^{+12}_{-10}$ & $160^{+13}_{-12}$ & $56^{+7}_{-6}$\\
\hline
500 & $97^{+9}_{-7}$ & $122^{+10}_{-9}$ & $30^{+4}_{-3}$\\
\hline
1000 & $50^{+5}_{-4}$ & $68^{+7}_{-5}$ &$10^{+1}_{-1}$ \\
\hline
\end{tabular}
\caption{\label{table}Bounds on $\Lambda$ at $95\%$ CL from the CMS $5.0 \, \fb^{-1}$ search for jets with $\slashed{E}_T$ including  loop-level processes in the analysis. The quoted errors reflect the scale uncertainties. See text for details.}
\end{table}

The scales involved in  $j + \slashed{E}_T$ production (i.e.\ the $p_T$ and the DM mass) are, however, not necessarily small compared to the top-quark mass, which implies that the infinite mass limit employed to obtain (\ref{eq:gluons}) is not a good approximation~\cite{Baur:1989cm}. Numerically, we find that applying the $m_t \to \infty$ limit overestimates the monojet cross sections by a factor of around  3 for small DM mass and that the quality of the approximation rapidly degrades with increasing $p_T$ cut and DM mass, resulting  in errors of up to a factor of 40. These numbers imply that the corresponding limits on $\Lambda$ would be too strong by  a factor of $1.2$ ($1.9$) in the best (worst) case. This  clearly shows that one cannot  use  $\mathcal{O}^\psi_{sg}$ (or analogous operators) to infer faithful bounds on the quark-DM contact operators.

To assess the theoretical errors in our analysis, we have studied the  scale ambiguities  of our results.  We find the scale $\mu$ which determines $\alpha_s (\mu)$ dynamically, i.e.\ we define $\mu = \xi  \hspace{0.5mm} p_T = \mu_ R = \mu_F$ and evaluate it on an event-by-event basis. Following common practice, we have varied $\xi$ in the range $[\frac{1}{2}, 2]$. We find that the predictions of the tree-level and loop-level monojet cross sections calculated in this way vary by around $\pm 20 \%$ and $\pm 50 \%$, respectively. The resulting uncertainties on $\Lambda$ stay below $\pm 2\,  {\rm GeV}$ and $\pm 15 \, {\rm GeV}$.  They are  given in  Tab.~\ref{table} and reflected by the width of the coloured bands in Fig.~\ref{Fig2}.

The pronounced scale ambiguity of the loop-level result indicates that next-to-leading order (NLO) corrections might be large. Indeed,  the $K$-factor representing the ratio between the NLO and LO cross sections of $p + p \to H + j$ evaluated in the $m_t \to \infty$  limit, varies between 1.2 and 1.8  depending on the kinematic region and choice of parton densities~\cite{deFlorian:1999zd,Ravindran:2002dc,Glosser:2002gm}. Using {\tt MCFM}, we explicitly verified that the same range of $K$-factors also apply to the monojet signal, although the imposed $p_T$ cut is significantly higher in this case than in the case of Higgs $+$ jet production. 

Another important and related issue is the relevance of events with a second high-$p_T$ jet. Such events are allowed in the CMS analysis, as long as the two jets are not back-to-back. To estimate the cross section for events with two jets and $\slashed{E}_T$, we use the cross section for $H + 2j$ implemented into MCFM in the limit $m_t \rightarrow \infty$~\cite{Campbell:2006xx,Campbell:2010cz}. This limit is known to work better in the case of  $H + 2j$ than for $H + j$~\cite{DelDuca:2001fn}. We find that the dominant contribution to two-jet events arises from processes that resemble monojet events but have an additional gluon in the final state. Allowing a second jet with large $p_T$ increases the total cross section by about a factor of 2, consistent with the more accurate simulations from CMS~\cite{Chatrchyan:2012pa}.

The inclusion of NLO corrections and of two-jet events is hence expected to strengthen the bounds on $\Lambda$, possibly by as much as $25\%$. However, we expect this improvement to be somewhat smaller in the full calculation with resolved top-quark loops (in particular, for high $p_T$ cut and large DM mass), so we prefer not to include these effects in our results and give a conservative bound. Clearly, a more careful analysis of finite top-quark mass effects at ${\cal O} (\alpha_s^4)$ (along the lines of the recent Higgs $+$ jet study \cite{Harlander:2012hf}) is required to make a more quantitative statement. This study is left for future work.

Finally, we note that the values of $\Lambda$ that we can exclude with the current data are low compared to typical LHC energies. To discuss the validity of the effective field theory~(EFT) approach, let us consider the  simplest ultraviolet (UV) completion, where~(\ref{eq:Opsis}) arises from the full theory~(\ref{eq:calLs}) after integrating out the scalar field $\Phi$. We anticipate that the new scalar or pseudoscalar mediator may be produced on-shell in $pp$ collisions, unless the couplings $g_q$ and $g_\psi$ in (\ref{eq:calLs}) are large (see~\cite{Shoemaker:2011vi,Fox:2011pm,Fox:2012ee}). In this case the analysis becomes more model-dependent, because the predictions now depend on $g_q$ and  $g_\psi$ as well as the mass  $M_\Phi$ and the decay width $\Gamma_\Phi$ of the mediator. 

The main features of the full theory calculation are captured by including a Breit-Wigner propagator for $\Phi$ in the monojet cross section~\cite{Fox:2011pm,Frandsen:2012rk}. In particular, the $j + \slashed{E}_T$ signal is enhanced relative to the EFT result if the $s$-channel mediator can be produced on-shell. In this case, using an EFT gives a lower bound on the expected monojet cross section. Only if the propagator is forced to be off-shell (because $M_\Phi \lesssim 2 m_\psi$) will the full theory predictions be reduced compared to the EFT. More precisely, for the $\slashed{E}_T$ and $p_T$ cuts that we employ and $m_\psi = 100 \, {\rm GeV}$,  we find that the EFT calculation underestimates the full theory cross section if $M_\Phi \gtrsim 280 \, {\rm GeV}$ $(M_\Phi \gtrsim 420 \, {\rm  GeV})$ for a decay width $\Gamma_\Phi = M_\Phi /( 8 \pi)$ ($\Gamma_\Phi = M_\Phi/ 3$). Such values of $M_\Phi$ are perfectly viable (in the sense that the couplings $g_q$ and $g_\psi$ are perturbative) given our bounds on $\Lambda_s$ and the relation $M_\Phi^2 = g_q \hspace{0.25mm} g_\psi \hspace{0.5mm} \Lambda_s^{3}/v$. For lighter mediators, on the other hand, the full theory including the Breit-Wigner propagator predicts an essentially constant cross section. The EFT calculation therefore overestimates the result and can no longer be used to give a conservative estimate of the monojet production cross section. 

In summary, as long as we focus on low-mass DM particles, which are best constrained by LHC monojet searches, the EFT is a good approximation that enables us to calculate lower bounds on the $j + \slashed{E}_T$ cross section. We find that the ratio between loop-level and tree-level cross sections calculated in the full theory is largely insensitive to the precise values of $g_q$, $g_\psi$, $M_\Phi$ and $\Gamma_\Phi$. Including the contributions of virtual top-quark loops gives an enhancement of more than two orders of magnitude irrespectively of whether the computation is done in the effective or full theory.  
 
\section{Bounds from relic density and direct searches}
\label{sec:conclusions}

The suppression scale $\Lambda$ which can be constrained through monojet searches also enters in the formula for the DM annihilation cross section and the DM-proton scattering cross section. Consequently, limits on $\Lambda$ can also be inferred from measurements of the relic density and DM direct detection experiments. Here we compare these results with the bounds derived in the previous section. 

Two different annihilation channels contribute to the total annihilation cross section. Tree-level annihilation into quarks will be dominant for $m_\text{DM} > m_t$, while annihilation into gluons via heavy-quark loops can give a relevant contribution for lower DM masses. At leading order in the DM velocity $v$, the annihilation cross sections into quarks for the three operators introduced in Sec.~\ref{Sect2} are given by (see e.g.~\cite{MarchRussell:2012hi})
\begin{align}
\big (\sigma^\psi_{s} v \big)_{\bar{\psi}\psi\rightarrow \bar{q} q } & = 
\frac{3 v^2 m_\text{DM}^4}{8 \pi \Lambda^6_s} \sum_q z_q \left(1 - z_q\right)^{3/2} \, ,\nonumber \\
\big (\sigma^\psi_{p} v\big)_{\bar{\psi}\psi\rightarrow \bar{q} q} & = 
\frac{3 m_\text{DM}^4}{2 \pi \Lambda^6_p} \sum_q z_q \left(1 - z_q\right)^{1/2} \, , \\
\big(\sigma^\phi_{s} v\big)_{\phi \phi \rightarrow \bar{q}  q} & = 
\frac{3 m_\text{DM}^2}{4 \pi \Lambda^4_s} \sum_q z_q \left(1 - z_q\right)^{3/2} \, , \nonumber
\end{align}
where $z_q = m_q^2 / m_\text{DM}^2$ and the sum includes the quarks with $z_q < 1$. The corresponding annihilation cross sections into gluons take the form (see e.g.~\cite{Chu:2012qy})
\begin{align}
 \big(\sigma^\psi_{s} v\big)_{\bar{\psi}\psi\rightarrow gg} & = \frac{v^2 \alpha_s^2  \hspace{0.25mm} m_\text{DM}^4}{18 \pi^3 \Lambda^6_s} \sum_q |F_s(z_q)|^2 \, , \nonumber \\ 
\label{eq:gluonannihilation}
\big (\sigma^\psi_{p} v\big)_{\bar{\psi}\psi\rightarrow gg} &  = \frac{\alpha_s^2 \hspace{0.25mm} m_\text{DM}^4}{2 \pi^3 \Lambda^6_p} \sum_q |F_p(z_q)|^2 \, ,\\
\big(\sigma^\phi_{s} v\big)_{\phi \phi \rightarrow gg} &  = \frac{\alpha_s^2  \hspace{0.25mm}  m_\text{DM}^2}{9 \pi^3 \Lambda^4_s} \sum_q |F_s(z_q)|^2 \, ,\nonumber
\end{align}
where the sum is over \emph{all} quarks, including those with  $z_q > 1$, and
 \begin{equation}
 \begin{split}
F_s(z) & = \frac{3 z}{2}   \left [ 1+ (1-z) \arctan^2 \left (\frac{1}{\sqrt{z-1}}\right) \right ] \,, \\[1mm]
F_p(z)  & = z \arctan^2 \left(\frac{1}{\sqrt{z-1}}\right) \,,
\end{split}
\end{equation}
are the well-known form factors for fermion loops~\cite{Djouadi:2005gi}. These form factors are normalised so that $F_s (\infty) = F_p (\infty) = 1$ in the limit of infinitely heavy quarks.

Using the expansion  $\sigma v = a + b v^2 + {\cal O} (v^4)$ the DM relic density $\Omega_{\mathrm{DM}}$ after freeze-out is given by~\cite{Kolb:1990vq}
\begin{equation}
\Omega_{\mathrm{DM}}h^2\simeq \frac{1.07\times10^9 }{\mathrm{GeV}} \frac{x_f}{M_{\mathrm{Pl}}\hspace{0.25mm}\sqrt{g_\ast} \left (a+\frac{3b}{x_f} \right)} \;,
\label{Om}
\end{equation}
where $g_\ast$ is the number of relativistic degrees of freedom at the freeze-out temperature $T_f$ and $x_f=m_\text{DM}/T_f$. If we assume that the interactions between the DM particle and SM quarks are the dominant ones at freeze-out, we can determine $\Lambda$ as a function of $m_{\rm DM}$ by the requirement that the observed DM relic abundance $\Omega_{\mathrm{DM}}h^2 \,  \simeq \, 0.1109$ \cite{Jarosik:2010iu} is reproduced. 

The values of $\Lambda$ which give the correct relic density are indicated by the black dashed curves in Fig.~\ref{Fig2}. Larger values of $\Lambda$ corresponds to DM overproduction in the early universe, smaller values imply underproduction. While the former is experimentally excluded, the latter is acceptable, assuming that the particle considered here accounts for only a fraction of the total DM abundance.  From the intersections of the monojet bounds and the relic density  constraints, we obtain the following lower bounds on the DM mass: 
\begin{equation} \label{eq:cons}
\begin{aligned}
\mathcal{O}^\psi_s: \quad & m_\psi > 129 \; (149) \; \text{GeV}~, \\
\mathcal{O}^\psi_p: \quad & m_\psi > 44 \; (55) \; \text{GeV}~, \\
\mathcal{O}^\phi_s: \quad & m_\phi > 1.4 \; (1.5) \; \text{GeV}~.
\end{aligned}
\end{equation}
 Here the values in brackets apply if DM is a Majorana fermion or a real scalar instead of  a Dirac fermion or a complex  scalar. It should be noted that whilst large regions of parameter space are excluded due to overproduction of DM, these bounds can be ameliorated if the DM has large annihilation cross sections to other SM states or (in particular) new hidden sector states. Such additional annihilation channels can in principle significantly reduce the tension between relic density constraints and direct detection experiments. Conversely, the correct DM relic density can still be obtained in principle if the DM is underproduced. For instance, if the hidden sector carries an particle-antiparticle asymmetry (similar to the baryon asymmetry) then this necessarily leads to a larger relic density compared to conventional (symmetric) DM  (see e.g.~\cite{MarchRussell:2012hi}). The constraints (\ref{eq:cons}) therefore  provide conservative lower bounds on such models of asymmetric DM.

\begin{figure}
\centering
 \includegraphics[width=0.7\columnwidth]{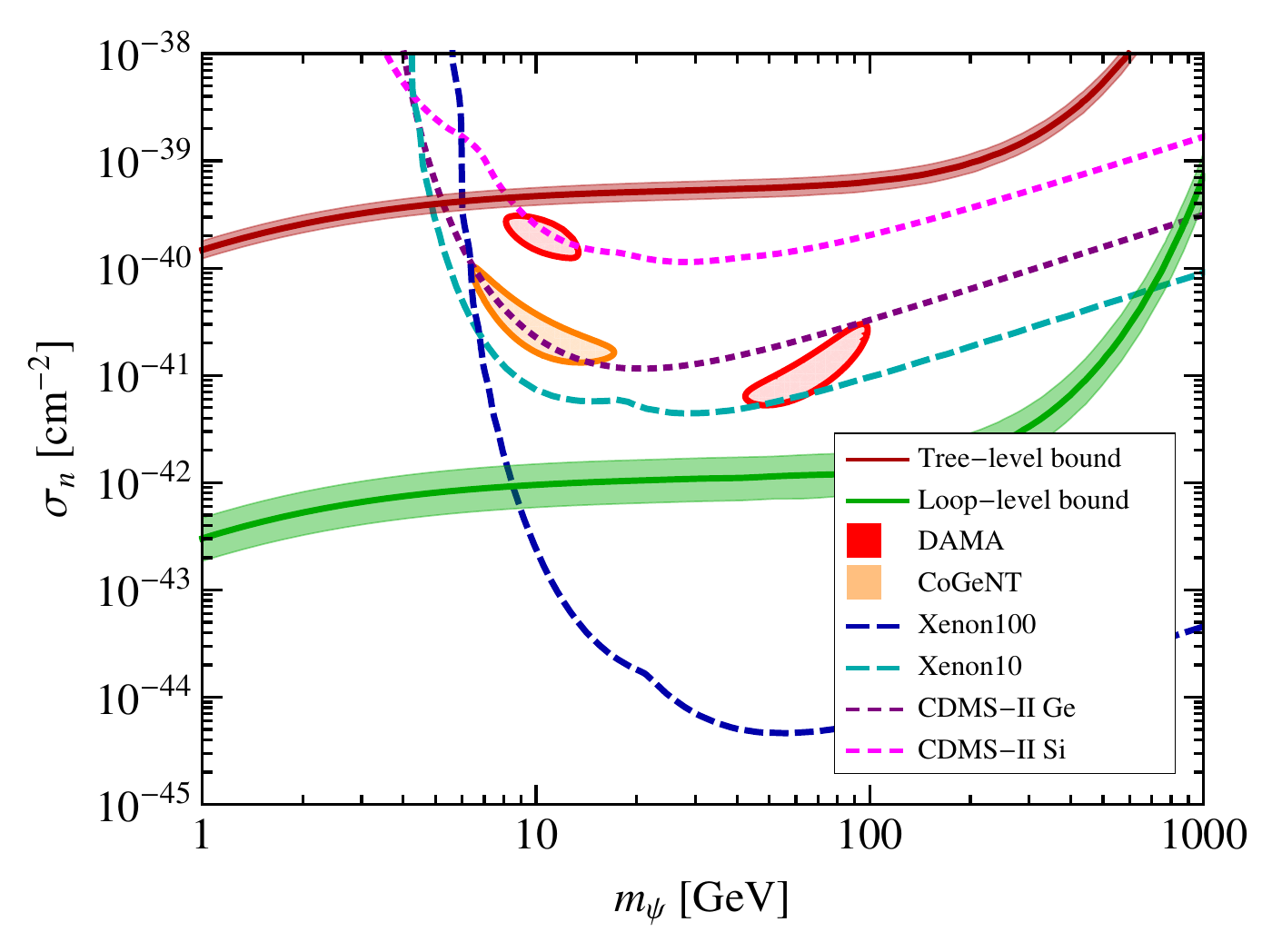}
 \caption{\label{Fig3} LHC monojet bounds on the DM-proton cross section for the operator $\mathcal{O}^\psi_s$ at tree-level (red) and loop-level (green) compared to various results from DM direct detection experiments. While the tree-level monojet bound is too weak to constrain the parameter regions favoured by DAMA and CoGeNT, the loop-level bound clearly excludes these regions.} 
\end{figure}

For the scalar operators $\mathcal{O}^{\psi}_s$ and $\mathcal{O}^{\phi}_s$, the DM direct detection cross section is given by
\begin{equation} \label{eq:sigmap}
\sigma^\psi_p = \frac{\mu_p^2 \hspace{0.5mm} m_p^2}{\pi} \frac{f^2}{\Lambda^6} \;, \qquad
\sigma^\phi_p  = \frac{\mu_p^2 \hspace{0.5mm} m_p^2}{\pi} \frac{f^2}{\Lambda^4 \hspace{0.5mm} m_\phi^2} \;,
\end{equation}
where $m_p$ is the proton mass, $\mu_p$ is the DM-proton reduced mass and $f \simeq0.35$ is the scalar form factor of the proton~\cite{Beltran:2008xg}. The scattering cross sections for Majorana or real scalar DM are larger by a factor of 4. Utilizing the formulas (\ref{eq:sigmap}), the monojet bounds on $\Lambda$ can be translated into limits on $\sigma_p$, which in turn can be compared to the exclusion limits obtained from XENON100~\cite{Aprile:2012nq}, XENON10~\cite{Angle:2011th} and CDMS-II~\cite{Ahmed:2010wy, Akerib:2010pv} as well as to the best-fit regions obtained from DAMA~\cite{Bernabei:2010mq} and CoGeNT~\cite{Aalseth:2011wp} \footnote{To obtain the CoGeNT best-fit region, we have subtracted surface events from the total event rate as discussed in~\cite{Kelso:2011gd}.}.

Our final results are shown in Fig.~\ref{Fig3} for the case where the DM particle is a Dirac fermion. All shown bounds and best-fit regions represent 95\% CL ranges. For large values of $m_\psi$, as indicated by the relic density constraints, direct detection experiments give stronger bounds than monojet searches. For $m_\psi \simeq10 \, {\rm GeV}$, the bounds become comparable, while below this value the bounds from LHC searches are far superior. We observe that the inclusion of loop-level processes gives a pertinent improvement of the monojet bounds, in particular because it excludes the possibility that the CoGeNT excess or the DAMA modulation arise from the interactions of a heavy scalar mediator.

For the pseudoscalar operators the DM direct detection cross section is spin-dependent and suppressed by $q^4 / m_p^4$, so that no relevant bounds on $\Lambda$ can be obtained from direct detection experiments. Consequently, the bounds shown in the central panel of Fig.~\ref{Fig2} are presently the strongest limit on the new-physics scale $\Lambda$. It is evident from the figure that including one-loop contributions improves the bound on $m_\psi$ inferred from the relic abundance by a factor of approximately 15.

\section{Conclusions}
\label{sec:summary}

While collider bounds on DM-quark contact operators with Yukawa-like couplings are relatively weak when only tree-level processes are considered, much stronger bounds can be obtained by including  heavy-quark loops in the analysis. In this letter, we used the recent CMS $5.0 \, \fb^{-1}$ and ATLAS $4.7 \, \fb^{-1}$ searches for jets with $\slashed{E}_T$ to obtain the  strongest collider limits on mass-dependent DM-quark scalar and pseudoscalar contact operators. Given that the LHC high-$p_T$ experiments are rapidly accumulating luminosity, constraints on all contact operators involving DM particles with masses below the electroweak scale will improve significantly in the near future. The methods outlined here will be important for further advances in constraining the parameter space of DM-quark interactions with mass-dependent couplings, and we are looking forward to their implementation in future LHC analyses.

\nocite{Kelso:2011gd}

\section*{Acknowledgements}
We are grateful to Keith~Ellis, Mads~Frandsen, Robert~Harlander, John~March-Russell, Michael~Rauch, Raoul~Rontsch, Subir~Sarkar, Kai~Schmidt-Hoberg, Alex~Tapper, Stephen West, Steven~Worm and Giulia Zanderighi for useful discussions.  UH acknowledges travel  support from the UNILHC network (PITN-GA-2009-237920). FK is supported by the Studienstiftung des Deutschen Volkes. JU is grateful for partial support from the Esson Bequest, Mathematical Institute, Oxford and the Vice-Chancellors' Fund, Oxford.

\providecommand{\href}[2]{#2}\begingroup\raggedright\endgroup

\end{document}